\documentclass[lineno]{jfm}

\usepackage{graphicx}
\usepackage{epstopdf, epsfig}
\usepackage{subfigure}
\usepackage{changes}
\usepackage{amsmath}
\usepackage{siunitx}

\usepackage{newtxtext}
\usepackage{newtxmath}
\usepackage{natbib}
\usepackage{hyperref}
\hypersetup{
    colorlinks = true,
    urlcolor   = blue,
    citecolor  = black,
}

\newcommand{\RomanNumeralCaps}[1]

\newcommand{\exsix}{\includegraphics[scale=0.02]{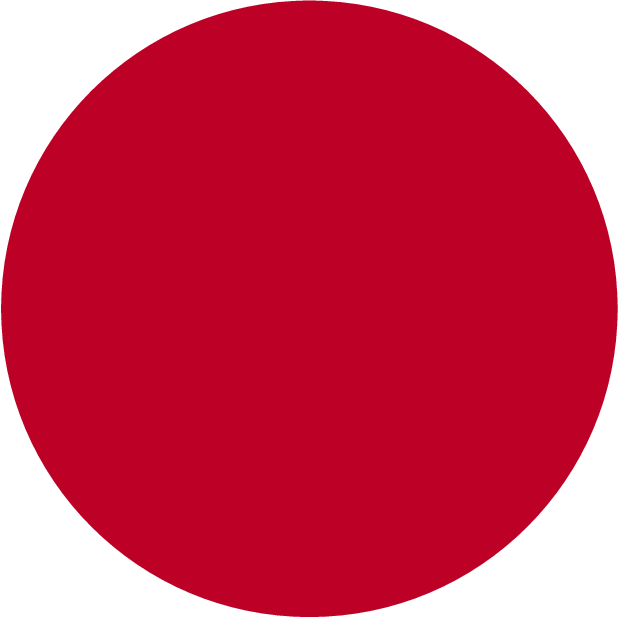}}
\newcommand{\exseven}{\includegraphics[scale=0.02]{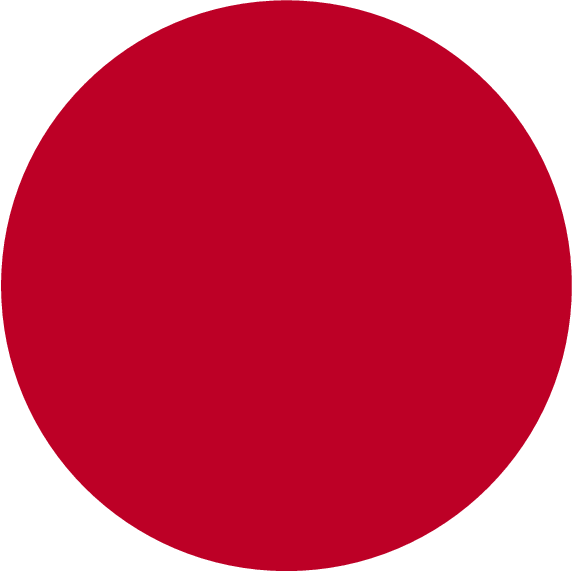}}
\newcommand{\exeight}{\includegraphics[scale=0.02]{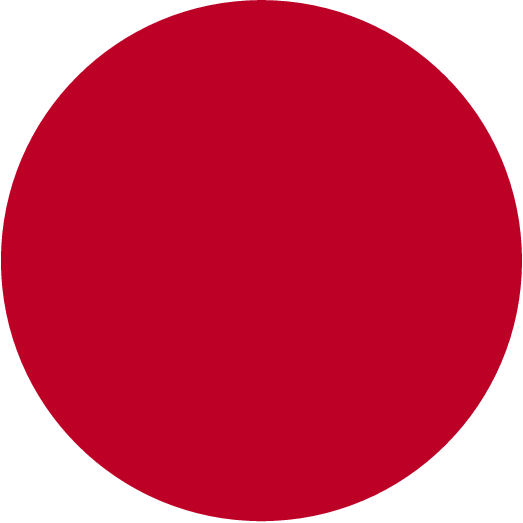}}
\newcommand{\exnine}{\includegraphics[scale=0.02]{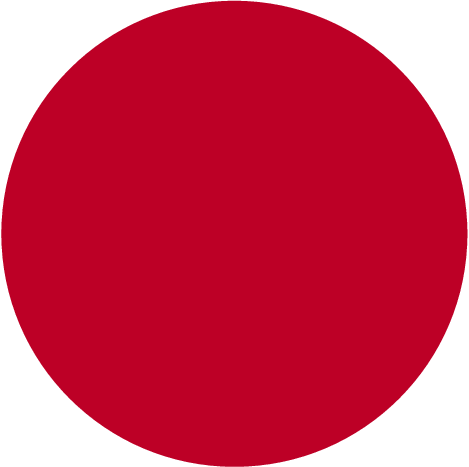}}
\newcommand{\exten}{\includegraphics[scale=0.02]{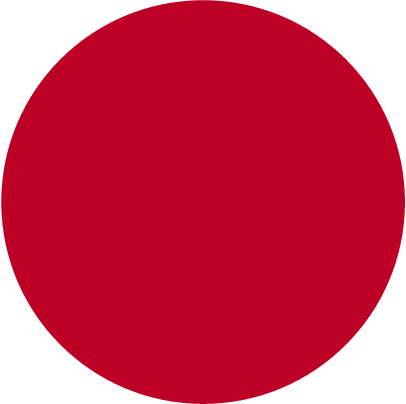}}
\newcommand{\exoone}{\includegraphics[scale=0.02]{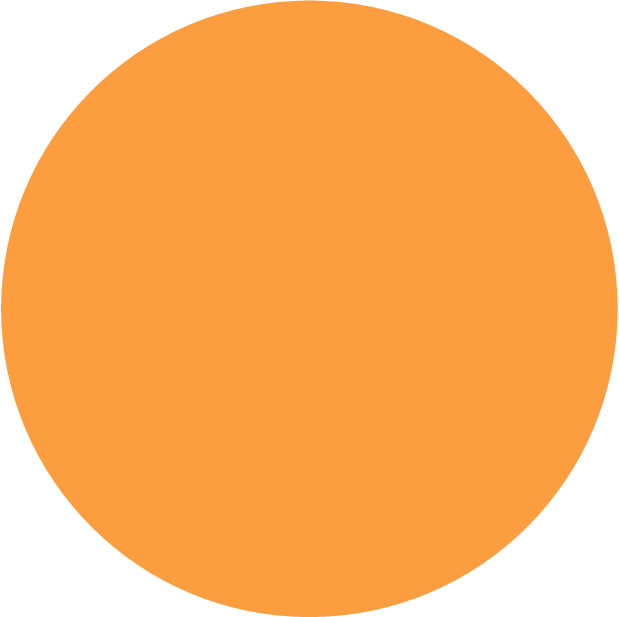}}
\newcommand{\exotwo}{\includegraphics[scale=0.02]{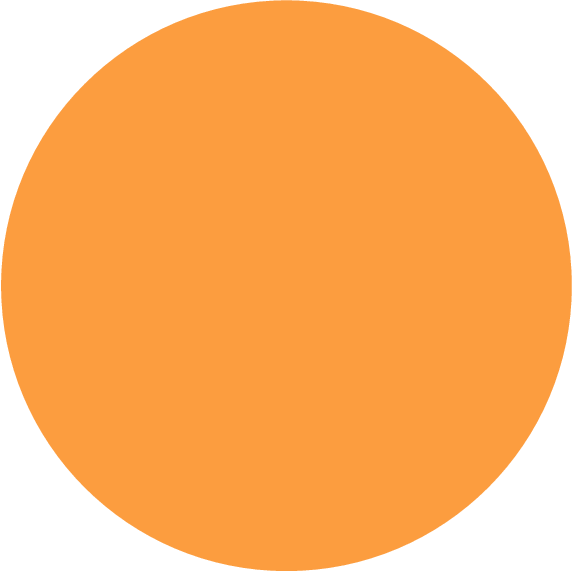}}
\newcommand{\exothree}{\includegraphics[scale=0.02]{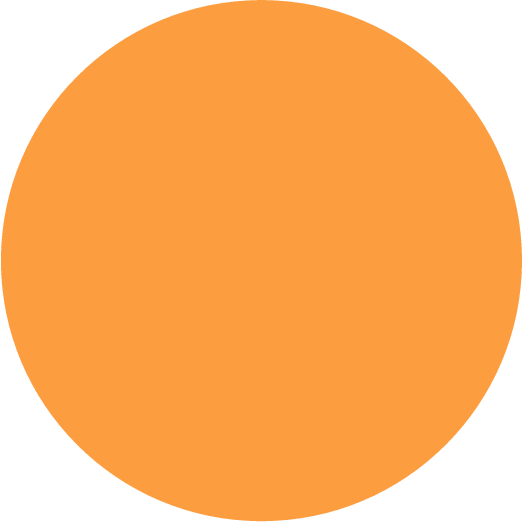}}
\newcommand{\exofour}{\includegraphics[scale=0.02]{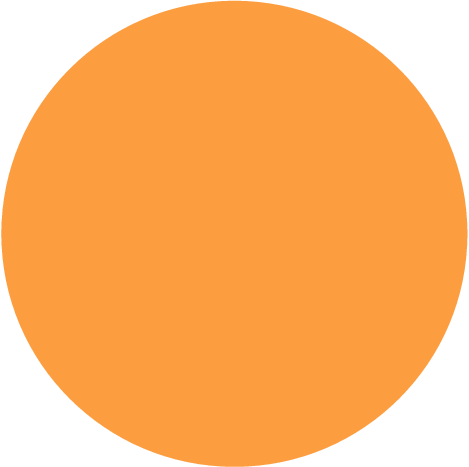}}
\newcommand{\exofive}{\includegraphics[scale=0.02]{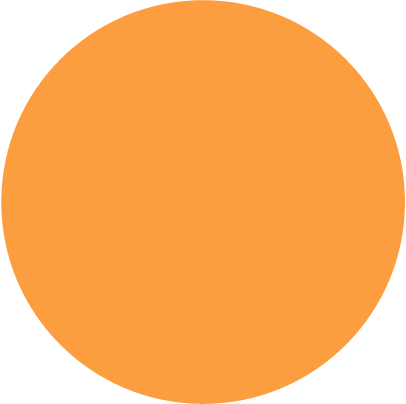}}
\newcommand{\exosix}{\includegraphics[scale=0.02]{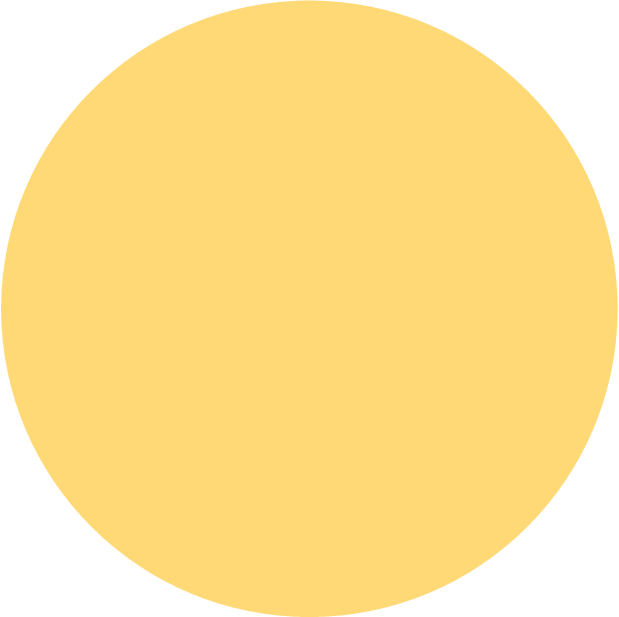}}
\newcommand{\exoseven}{\includegraphics[scale=0.02]{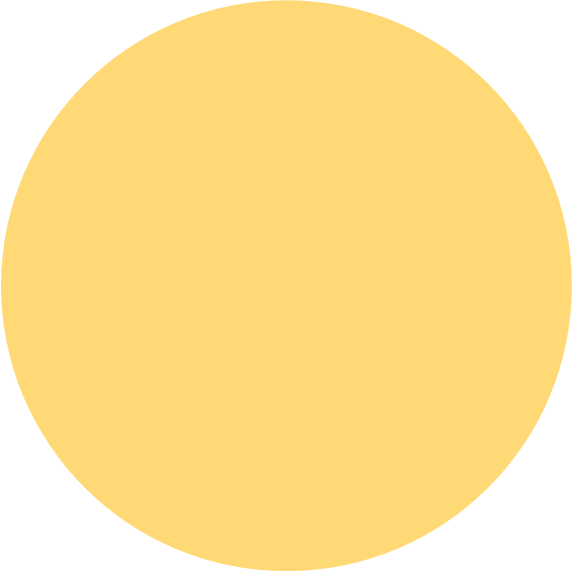}}
\newcommand{\exoeight}{\includegraphics[scale=0.02]{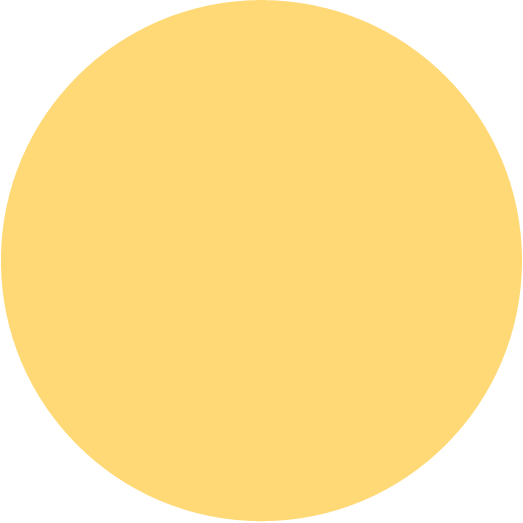}}
\newcommand{\exonine}{\includegraphics[scale=0.02]{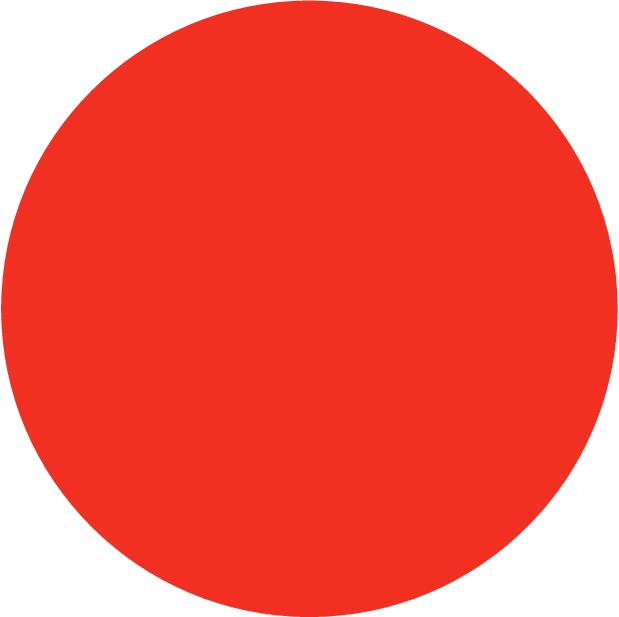}}
\newcommand{\extwenty}{\includegraphics[scale=0.02]{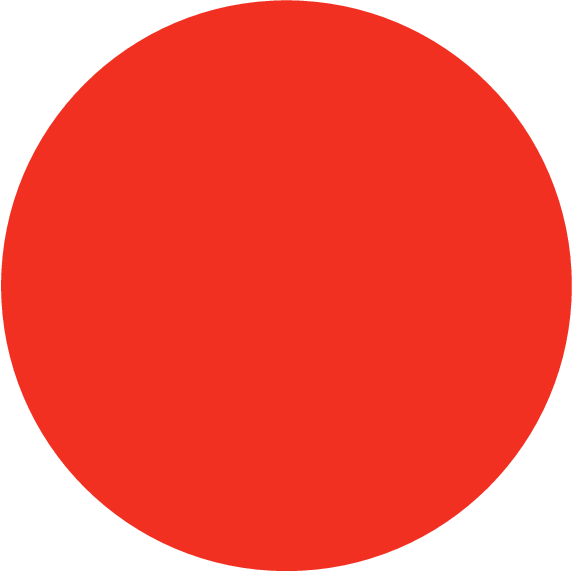}}
\newcommand{\extone}{\includegraphics[scale=0.02]{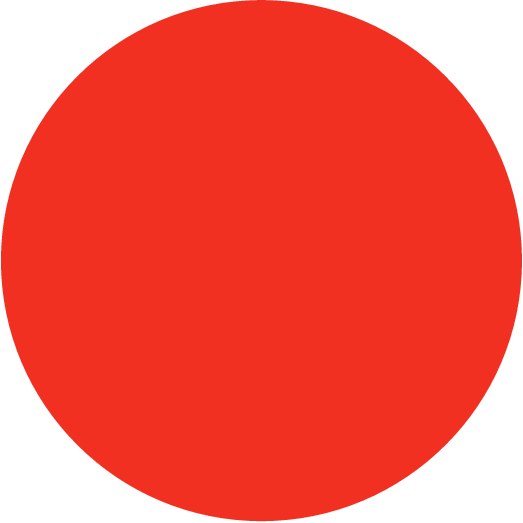}}
\newcommand{\exttwo}{\includegraphics[scale=0.02]{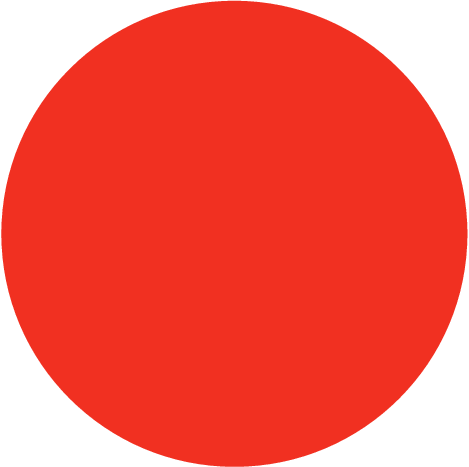}}
\newcommand{\extthree}{\includegraphics[scale=0.02]{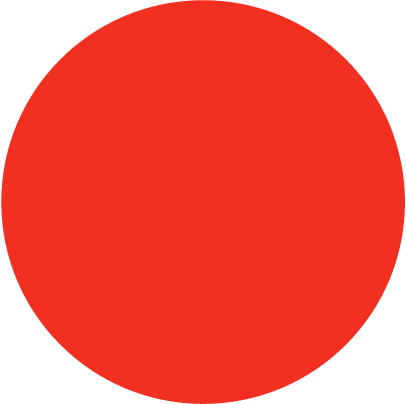}}

\title{Drag reduction utilizing a wall-attached ferrofluid film in turbulent channel flow,\nolinenumbers}

\author{\nolinenumbers Marius M. Neamtu-Halic\aff{1,2}
  \corresp{\email{nemarius@ethz.ch}},
  Markus Holzner\aff{1,3,4}
 \and Laura M. Stancanelli\aff{5}}

\affiliation{\nolinenumbers\aff{1}Swiss Federal Institute of Forest, Snow and Landscape Research WSL, 8903 Birmensdorf, Switzerland
\aff{2}Institute of Environmental Engineering, ETH Zürich, 8039 Zürich, Switzerland
\aff{3}Swiss Federal Institute of Aquatic Science and Technology Eawag, 8600 Dübendorf, Switzerland
\aff{4}Institute of Hydraulic Engineering and River Research, University of Natural Resources and Life Sciences, BOKU Wien, 
1190 Vienna, Austria
\aff{5}Department of Hydraulic Engineering, Delft University of Technology, TU Delft, 2628CN Delft, The Netherlands}

\begin{document}

\maketitle
\nolinenumbers
\begin{abstract}

This study explores the application of a wall-attached ferrofluid film to decrease skin friction drag in turbulent channel flow. We conduct experiments using water as a working fluid in a turbulent channel flow setup, where one wall is coated with a ferrofluid layer held in place by external permanent magnets. Depending on the flow conditions, the interface between the two fluids is observed to form unstable travelling waves. While ferrofluid coating has been previously employed in laminar and moderately turbulent flows to reduce drag by creating a slip condition at the fluid interface, its effectiveness in fully developed turbulent conditions, particularly when the interface exhibits instability, remains uncertain. Our primary objective is to assess the effectiveness of ferrofluid coating in reducing turbulent drag with particular focus on scenarios when the ferrofluid layer forms unstable waves. To achieve this, we measure flow velocity using two-dimensional particle tracking velocimetry (2D-PTV), and the interface contour between the fluids is determined using an interface tracking algorithm. Our results reveal the significant potential of ferrofluid coating for drag reduction, even in scenarios where the interface between the surrounding fluid and ferrofluid exhibits instability. In particular, waves with an amplitude significantly smaller than a viscous length scale positively contribute to drag reduction, while larger waves are detrimental, because of induced turbulent fluctuations. However, for the latter case, slip out-competes the extra turbulence so that drag is still reduced.

\end{abstract}


\section{Introduction}
\label{sec:Introduction}

Over the last decades, a considerable research effort has been dedicated to the reduction of drag in turbulent flows. This interest stems primarily from the goal of decreasing energy consumption in various applications, such as fluid transportation through pipes or the movement of a vessel through a fluid. 

Drag reduction techniques are commonly classified into passive and active, the later requiring external energy to be fed into the system \citep{gad1996modern,quadrio2004critical}.

Passive drag reduction techniques can be classified into two categories: additives, primarily polymers mixed into the fluid, and surface property modification techniques. While the former are employed in the turbulent regime only, the latter are especially efficient in laminar flows. Additives are used to reduce turbulence intensity or alter the apparent viscosity of the fluid \citep{lumley1969drag, white2008mechanics, zhang2021drag}. Virk's influential work in the 1970s demonstrated that the drag reduction achieved with polymer additives asymptotically approaches a maximum value \citep{virk1970ultimate}, the so-called Virk asymptote. On the other hand, surface property modification techniques focus on altering the characteristics of the wall. A typical example is the creation of micro- and nano-textures that trap microscopic gas pockets, resulting in the Cassie-Baxter state, which is a superhydrophobic surface \citep{liravi2020comprehensive}. Other examples include the use of riblets \citep{saranadhi2016sustained, costantini2018drag, dean2010shark} and the application of chemical surface coatings \citep{solomon2014drag, kim2020lubricant}. 

Initial observations of natural surfaces in contact with fluids, such as shark skin, prompted scientists to investigate the impact of surface texture on drag reduction. They discovered that regular riblets on solid surfaces can hinder the formation of coherent flow structures near the surface \citep{choi1993direct}. These vortical flow structures promote the transfer of momentum perpendicular to the surface, and their suppression dampened this transfer and hence drag \citep{kim1993propagation, dean2010shark}. Building upon these findings, recent active techniques have emerged that utilize cyclic wall movements perpendicular to the flow direction to control the formation of vortical structures near solid surfaces, aiming to reduce drag \citep{quadrio2004critical,quadrio2009streamwise,ricco2021review}. Remarkably, drag reduction of up to 45\% was achieved under optimal conditions, although the energy balance between the power saved through oscillation and the power expended to activate the wall motions was typically limited to 7\% at the most. Another established technique involves creating streamwise traveling waves of wall deformation. Depending on the maximum wall displacement and wave periods, this technique can result in either drag reduction or an increase in drag \citep{nakanishi2012relaminarization,nabae2020prediction}. For a comprehensive review of this method, we refer to the recent work by \citet{fukagata2024turbulent}.           

Recent investigations have delved into harnessing magnetically controllable fluids, particularly ferrofluids, to form a slim coating on the wall effectively establishing a slip boundary condition. Although this technique was initially developed for microfluidic devices \citep{dev2022fluid}, its feasibility in macroscopic flows has recently been demonstrated by \citet{sta2024phys} in laminar and moderately turbulent regimes. However, it is known that the ferrofluid interface can be subject to instability giving rise to traveling waves at high flow velocities, posing the question about their effectiveness at higher Reynolds numbers \citep{kogel2020calming,volkel2020measuring}. 

The aim of this work is to reveal the drag reduction efficiency of a ferrofluid coating when instabilities at the interface occur in the turbulent regime. Our goal is to systematically explore various turbulent flow regimes and magnetic field strengths to establish connections between interface wave characteristics and drag reduction. 

The structure of this paper is outlined as follows. Section \ref{sec:Methods} provides a detailed description of the methodology employed. Section \ref{sec:Results} presents the findings obtained from the study. Finally, the paper concludes with a summary and the main conclusions in section \ref{sec:Conclusions}.

\section{Methods}\label{sec:Methods}

\begin{figure}
  \centerline{\includegraphics[width=\textwidth]{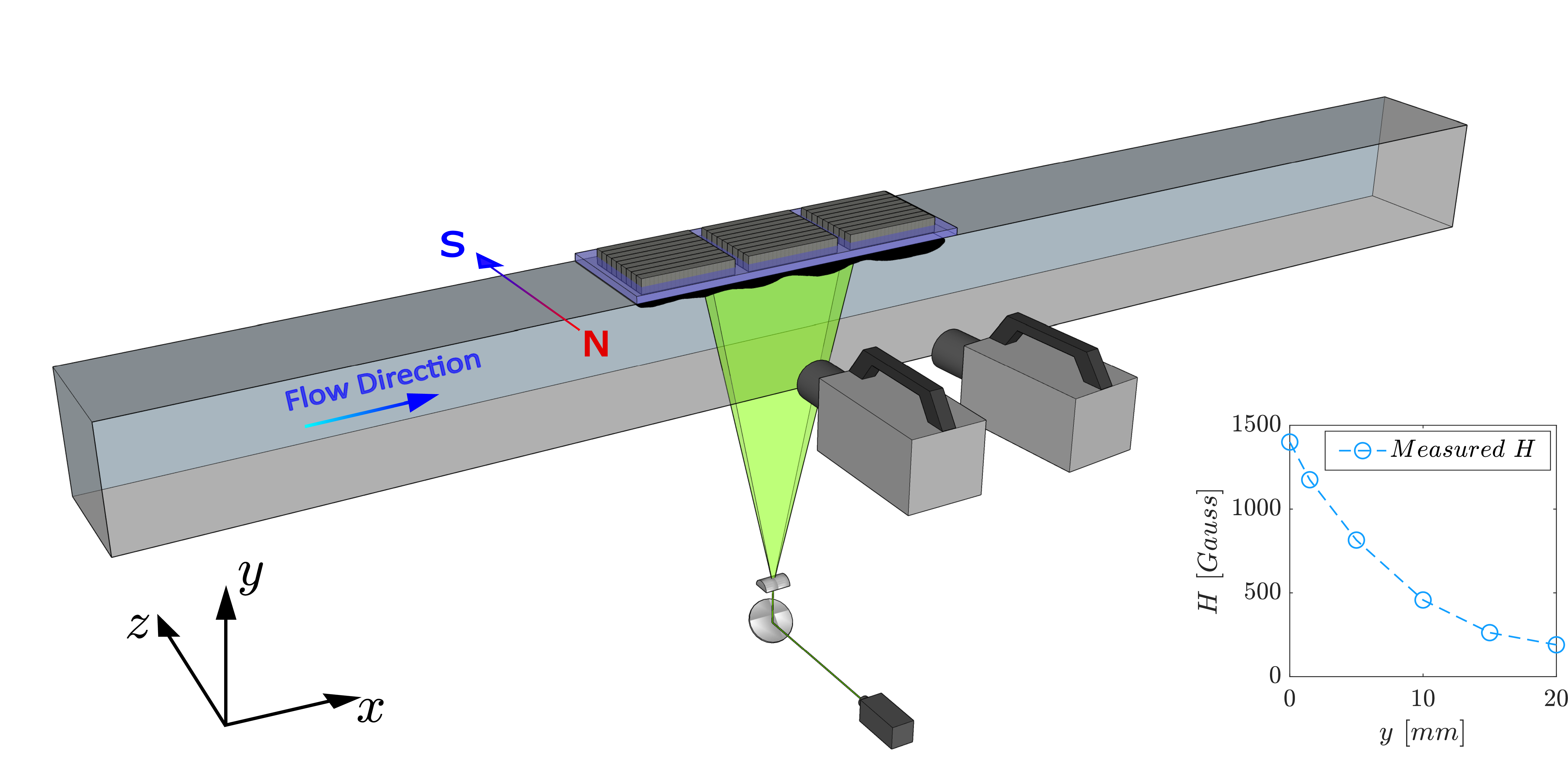}}
  \caption{Sketch of the setup and measured spanwise magnetic field intensity $H$ as a function of distance $y$ from the permanent magnets.}
\label{fig:fig1}
\end{figure}

\subsection{Experimental Setup}\label{subsec:Eperimental Setup}

In this work, experiments were conducted using a Plexiglas channel, with a square cross section of dimensions $25 \times 25 mm^2$ and $1000 mm$ in length.


A layer of ferrofluid was introduced at the top of the channel at a distance of $500 mm$ from the conduit inlet. The ferrofluid layer had a length of around $300mm$ and a height of approximately $3mm$. To hold the ferrofluid in place, 14 stacks of $9$ magnets (each with dimensions $3\times3\times20 mm^3$ from "Supermagnete" - Neodymium 45) were positioned at gap intervals of $2 mm$ in the $x$ (streamwise) direction covering the entire length of the ferrofluid layer. The arrangement is depicted schematically in Figure \ref{fig:fig1}.

The magnetic field generated by the magnets was tangentially oriented to the ferrofluid layer in the $z$ spanwise direction. The intensity of the magnetic field, denoted as $H$, was measured using a "PCE-MFM 4000" Gaussmeter. Figure \ref{fig:fig1} illustrates the tangential magnetic field intensity as a function of the distance from the magnets. As observed, the magnetic field exhibits an exponential decrease with distance $y$ in the vertical direction. Throughout this study, we varied the magnetic field intensity by adjusting the position of the magnet array relative to the top wall of the channel.  

A closed loop circulation of water through the channel was realized with the following specifications. Water was pumped out of a reservoir using a water pump model "LIVERANI EP-MINI 3/4" and directed to a water tower with fixed pressure head. From there, the water was directed through the Plexiglas channel and returned to the reservoir. Control over flow discharge and pressure inside the channel was achieved using two valves positioned upstream and downstream of the Plexiglas channel, respectively. The flow rate was measured by a flow meter (type "SIEMENS SITRANS FM MAG 1100") installed at the inlet of the channel.

High speed imaging was used to characterize the ferrofluid interface and the flow at a distance of $200mm$ away from the ferrofluid injection point, where the interface waves and the flow reached an equilibrium, i.e. conditions are homogeneous in streamwise direction. Flow measurements were performed using two-dimensional Particle Tracking Velocimetry (2D-PTV) in a vertical plane aligned with the flow direction and passing through the center of the channel (Figure \ref{fig:fig1}). For this purpose, the flow was seeded with neutrally buoyant polystyrene particles with a diameter of $11 \mu m$. To visualize the particles, a laser sheet was generated using a solid state 532nm green laser (model "MGL-FN-532"). A high-speed camera (Photron SA5), operating at a frequency of 4000 Hz, was employed to capture the movement of the particles within an observation window of approximately $25\times25 mm^2$. In our measurements, we estimate relative errors of $5\%$ for velocity measurements and $12\%$ for shear stresses \citep{sta2024phys}.

To analyze the motion of the interface, a second high-speed camera (operating at frequency in the range of $1000 - 3000Hz$) was employed, which covered a larger observation volume of approximately $60\times25 mm^2$. To identify and extract the interface, a median filter with a kernel size of approximately 11 pixels was first applied. Then, the interface was detected using a threshold on the gray value intensity. Based on this method, the local interface height is estimated with a relative error of $1\%$. 
Exemplary videos showing the motion of the interface are provided in the Supplementary Material. 

\subsection{Experiments}\label{subsec:Exps}

A total of 18 experiments were caried out, at four different magnetic field intensities $H$ in the range 140 - 460 Gs (see Table \ref{tab:tab1}). For each magnetic field intensity, five different flow rates were used varying the Reynolds numbers $Re=U \nu /(D-h)$ in the range 2500-12500, where $U$ is the space and time average streamwise flow velocity, $\nu$ is the water kinematic viscosity, 
$D$ the channel height and $h$ the ferrofliud layer height. For the lower magnetic field intensity $H=140Gs$, only three experiments were conducted as the ferrofluid layer was subject to detachment at the two highest Reynolds number cases.The experimental parameters are reported in Table \ref{tab:tab1}.

\begin{table}
  \begin{center}
\def~{\hphantom{0}}
  \begin{tabular}{lccccc}
      $Label$  & $Re[-]$   &   $H[Gs]$ & $h[mm]$ & $U[m/s]$ & $Symbol$\\[3pt]
       a.1   & 2740 & 460 & 2.95 & 0.11 & \exsix\\
       a.2   & 5080 & 460 & 3.04 & 0.20 & \exseven\\
       a.3  & 7430 & 460 & 3.04 & 0.30 & \exeight\\
       a.4   & 9850 & 460 & 3.09 & 0.40 & \exnine\\
       a.5 & 12380 & 460 & 3.11 & 0.50 & \exten\\
       b.1   & 2580 & 250 & 3.13 & 0.10 & \exonine\\
       b.2   & 5030 & 250 & 3.20 & 0.20 & \extwenty\\
       b.3  & 7200 & 250 & 3.25 & 0.30 & \extone\\
       b.4   & 9820 & 250 & 3.28 & 0.40 & \exttwo\\
       b.5 & 12230 & 250 & 3.38 & 0.50 & \extthree\\
       c.1   & 2780 & 170 & 2.87 & 0.10 & \exoone\\
       c.2   & 5270 & 170 & 3.14 & 0.21 & \exotwo\\
       c.3  & 7740 & 170 & 3.12 & 0.31 & \exothree\\
       c.4   & 10150 & 170 & 3.10 & 0.40 & \exofour\\
       c.5 & 12660 & 170 & 2.82 & 0.50 & \exofive\\
       d.1   & 3090 & 140 & 3.21 & 0.12 & \exosix\\
       d.2   & 5100 & 140 & 2.88 & 0.21 & \exoseven\\
       d.3  & 7300 & 140 & 3.23 & 0.30 & \exoeight\\
  \end{tabular}
  \caption{Overview of the experimental parameters, where $Re$ is the Reynolds number, $H$ the magnetic field intensity, $h$ the time-space average height of the ferrofluid layer, $U$ the time-space average streamwise velocity of ambient fluid. Letters ranging from "a" to "d" on the label represent the magnetic field intensity ($H$), while the numbers correspond to the Reynolds number. The color-code depicts the magnetic field intensity while the bullet size varies with $Re$ number.}
  \label{tab:tab1}
  \end{center}
\end{table}

\section{Results and Discussion}\label{sec:Results}

\subsection{Stability of the ferrofluid interface}\label{subsec:Interface Characterization}

\begin{figure}
  \centerline{\includegraphics[scale=0.20]{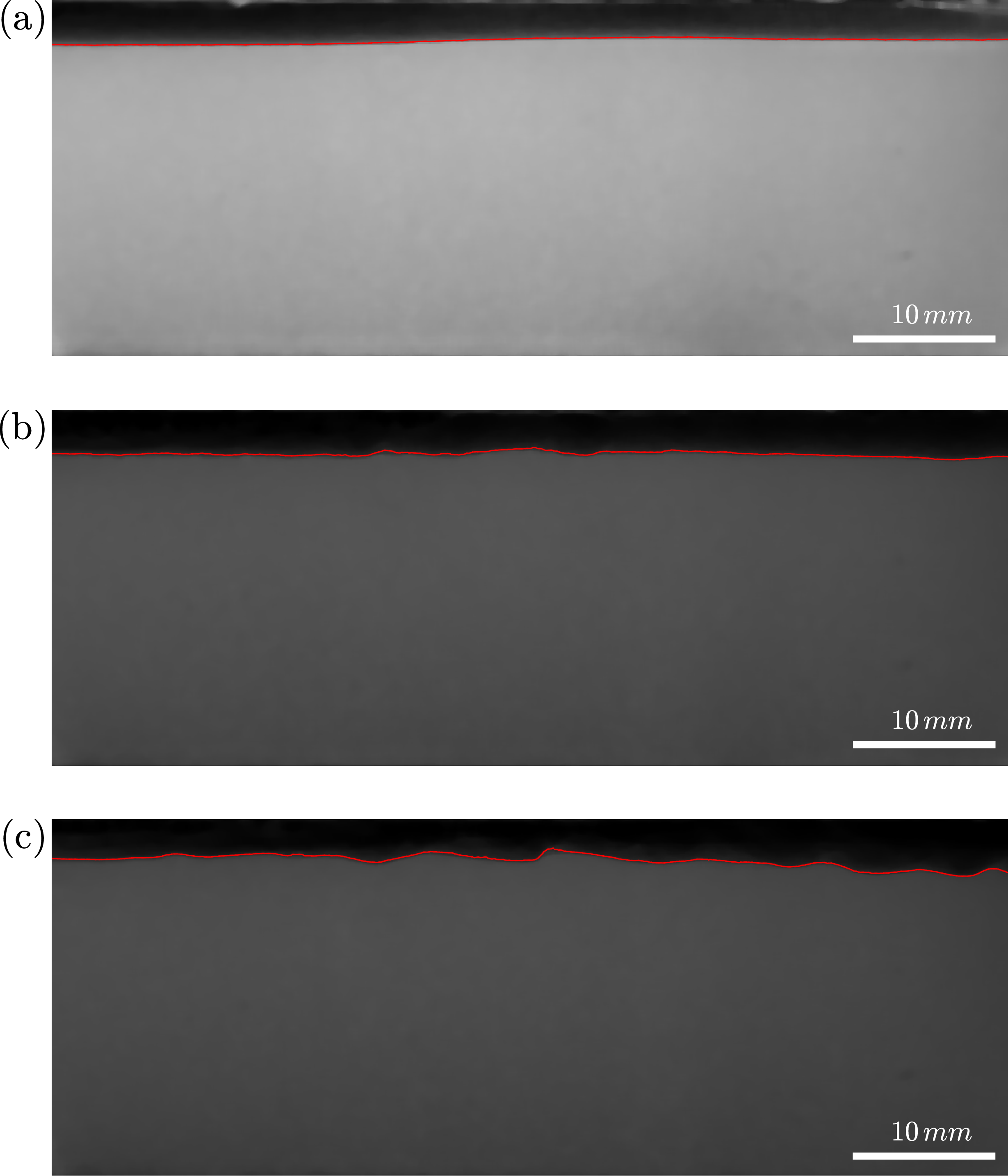}}
  \caption{Instantaneous snapshot of the ferrofluid interface at $H=170Gs$ for $Re=2780$, c.f., experiment c.1 (a), $Re=7740$ c.3 (b) and $Re=12660$ c.5 (c).}
\label{fig:fig2}
\end{figure}

As a first step in our analysis, we scrutinize the behavior of the wall coating, with a focus on the dynamics of the interface between ferrofluid and water. In Figure \ref{fig:fig2}, we show a snapshot of the interface at different Reynolds numbers and fixed magnetic field intensity ($H=170Gs$). When the Reynolds number is low (a), the interface appears rather flat. However, as the Reynolds number is raised, we observe the emergence of an interface instability, manifested in the form of traveling waves. 

To characterize the waves, we conduct an image post-processing analysis as follows. The interface height is denoted as $h(x,t)$, where $x$ represents the streamwise coordinate and $t$ denotes the time. The interface elevation relative to its mean is $\zeta(x,t) = h(x,t) - \overline{h(x)}$, where $\overline{h(x)}$ is the time-averaged height of the interface. To characterize the unstable waves, we identify the zero upcrossings of the time signal $\zeta(x_{0},t)$ at various locations $x=x_{0}$ (10 points along the horizontal coordinate with equal spacing of $\approx5.5mm$). Subsequently, we calculate the average wave amplitude, $a$, defined as the difference between the maximum and minimum of $\zeta(x_{0},t)$ between two adjacent zero upcrossings, and then average this value over all events and $x_{0}$ locations. Similarly, the wave period $T$ is computed as the average time span between zero upcrossings, and the wave frequency $\omega$ is defined as $\omega = 1/T$. Finally, to calculate the average wavelength $L$, we calculate the distance between the zero upcrossings of the spatial signal $\zeta(x,t_{0})$ at fixed times $t_{0}$ and average this measurement over all instances.

A synthesis of average wave characteristics across all experiments is outlined in Table \ref{tab:tab2}.

\begin{table}
  \begin{center}
 \def~{\hphantom{0}}
   \begin{tabular}{lccccc}
       $Label$  & $a[mm]$   &   $L[mm]$ & $\omega [1/s]$ & $Symbol$ & $Stability$\\[3pt]
        a.1   & 0.02 & 1.40 & 82.6 & \exsix & stable\\
        a.2   & 0.04 & 1.64 & 83.1 & \exseven & stable\\
        a.3  & 0.02 & 2.45 & 52.5 & \exeight & stable\\
        a.4   & 0.03 & 2.00 & 106.1 & \exnine & stable\\
       a.5 & 0.10 & 2.47 & 103.1 & \exten & unstable\\
        b.1   & 0.02 & 1.57 & 92.3 & \exonine & stable\\
        b.2   & 0.02 & 1.66 & 77.4 & \extwenty & stable\\
        b.3  & 0.03 & 1.72 & 119.8 & \extone & unstable\\
        b.4   & 0.18 & 2.62 & 87.7 & \exttwo & unstable\\
        b.5 & 0.16 & 3.91 & 68.7 & \extthree & unstable\\
        c.1   & 0.02 & 1.53 & 58.2 & \exoone & stable\\
        c.2   & 0.02 & 1.94 & 88.9 & \exotwo & unstable\\
        c.3  & 0.13 & 4.93 & 35.1 & \exothree & unstable\\
        c.4   & 0.46 & 6.28 & 30.1 & \exofour & unstable\\
        c.5 & 0.33 & 6.57 & 34.4 & \exofive & unstable\\
        d.1   & 0.07 & 1.70 & 65.4 & \exosix & stable\\
        d.2   & 0.04 & 2.01 & 100.9 & \exoseven & unstable\\
        d.3  & 0.42 & 7.11 & 26.3 & \exoeight & unstable\\
   \end{tabular}
   \caption{Overview of wave characteristics, where $a$ is the wave amplitude, $L$ the wave length and $\omega$ the wave time frequency. Letters ranging from "a" to "d" on the label represent the magnetic field intensity ($H$), while the numbers correspond to the Reynolds number. The color-code depicts the magnetic field intensity while the symbol size varies with $Re$ number.}
   \label{tab:tab2}
   \end{center}
 \end{table}

\begin{figure}
  \centerline{\includegraphics[scale=0.7]{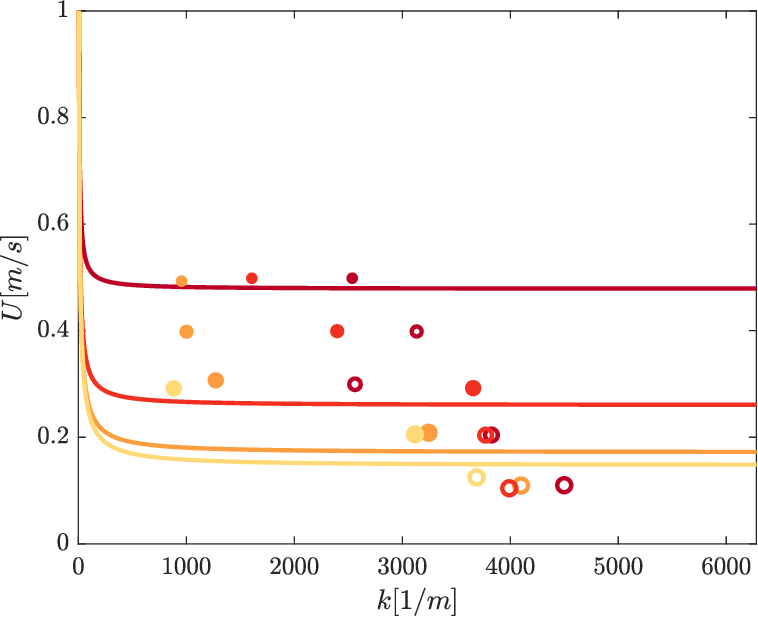}}
  \caption{Stability maps for the experiments in table \ref{tab:tab1}. The color-coding and bullet size are as specified in Table \ref{tab:tab1}. Empty symbols denote experiments featuring a stable interface, while filled symbols indicate cases characterized by an unstable interface.}
\label{fig:fig3}
\end{figure}

To contextualize the relative influence of the governing parameters, i.e. average water flow velocity $U$ and magnetic field intensity, we employ linear stability analysis. To this end, we compute the critical velocity curves, denoted as $U_{crit}$, which represent the average water velocity threshold for the onset of unstable travelling waves. These critical velocity curves are determined as a function of the magnetic field intensity $H$ and the wavenumber $k$ \citep{kogel2020calming, volkel2020measuring}, defined as:

\begin{equation}
\label{eq:Up}
    U_{crit}(k,H) = \sqrt{\frac{\rho_{f}+\rho_{w}}{\rho_{f}\rho_{w}}\left( k \gamma+\tilde{\mu}H^{2}+\frac{g(\rho_{f}-\rho_{w})}{k} \right)},   
\end{equation}

where $\rho_{f}$ is the ferrofluid density, $\rho_{w}$ is the ambient fluid density, $\gamma$ is the interface tension between the two fluids and $\tilde{\mu} = \mu_{0} \chi^2 /(\chi+2)$ is proportional to the vacuum permeability $\mu_{0}$ and depends on the susceptibility $\chi$ (the ferrofluid properties are given in Table \ref{tab:tab3}). Figure \ref{fig:fig3} displays the critical velocity neutral curves for our experiments, along with the measured values of $U$ (the bulk average flow velocity) and $k=2\pi /L$ (the average wave number of the interface). The neutral curves separate stable (below) from unstable (above) behavior. The increase in the intensity of the magnetic field $H$ is a stabilizing factor for the ferrofluid interface (shifting the neutral curve upwards). Conversely, an increase in the average flow velocity of the ambient fluid is destabilizing. Experiments with a given magnetic field intensity show that as $Re$ (represented by color-matched dots, ranging from larger to smaller) increases, there is a decrease in wave number ($k$).
Notably, when the magnetic field is at its strongest, $H=460Gs$, only one flow case exhibits instability ($Re=12380$, i.e., experiment a.5 in Table \ref{tab:tab1}). However, with a magnetic field intensity of $H=170Gs$, nearly all flow cases exhibit instability, except for one experiment (experiment c.1 in Table \ref{tab:tab1} corresponding to $Re=2780$). This is qualitatively consistent with Figure \ref{fig:fig2} where the interface is rather smooth for c.1, whereas for c.3 and c.5 it is corrugated. 

\begin{table}
  \begin{center}
 \def~{\hphantom{0}}
   \begin{tabular}{cc}
       Property  & Value\\[3pt]
        density $\rho_{f}\ [kg/m^3]$   & $1210$ \\
        viscosity $\mu_{f}\ [cSt]$   & $6$ \\
        magnetic susceptibility $\chi [T/(A/m)]$   & $0.32$ \\
        surface tension $\gamma [mN/m]$   & $43$ \\
   \end{tabular}
   \caption{Properties of ferrofluid EFH1 - "Ferrotec USA corporate".}
   \label{tab:tab3}
   \end{center}
 \end{table}

To statistically characterize the wave frequency spectra, we conduct a Fourier analysis of the interface elevation. This involves computing the temporal power spectral density of the interface elevation, $\zeta(x,t)$. The power spectra are then averaged over all locations to obtain a single average temporal power spectrum density for each experiment. The power spectrum computation utilizes Welch's method through the "pwhelch" function in MATLAB.

\begin{figure}
  \centerline{\includegraphics[scale=0.58]{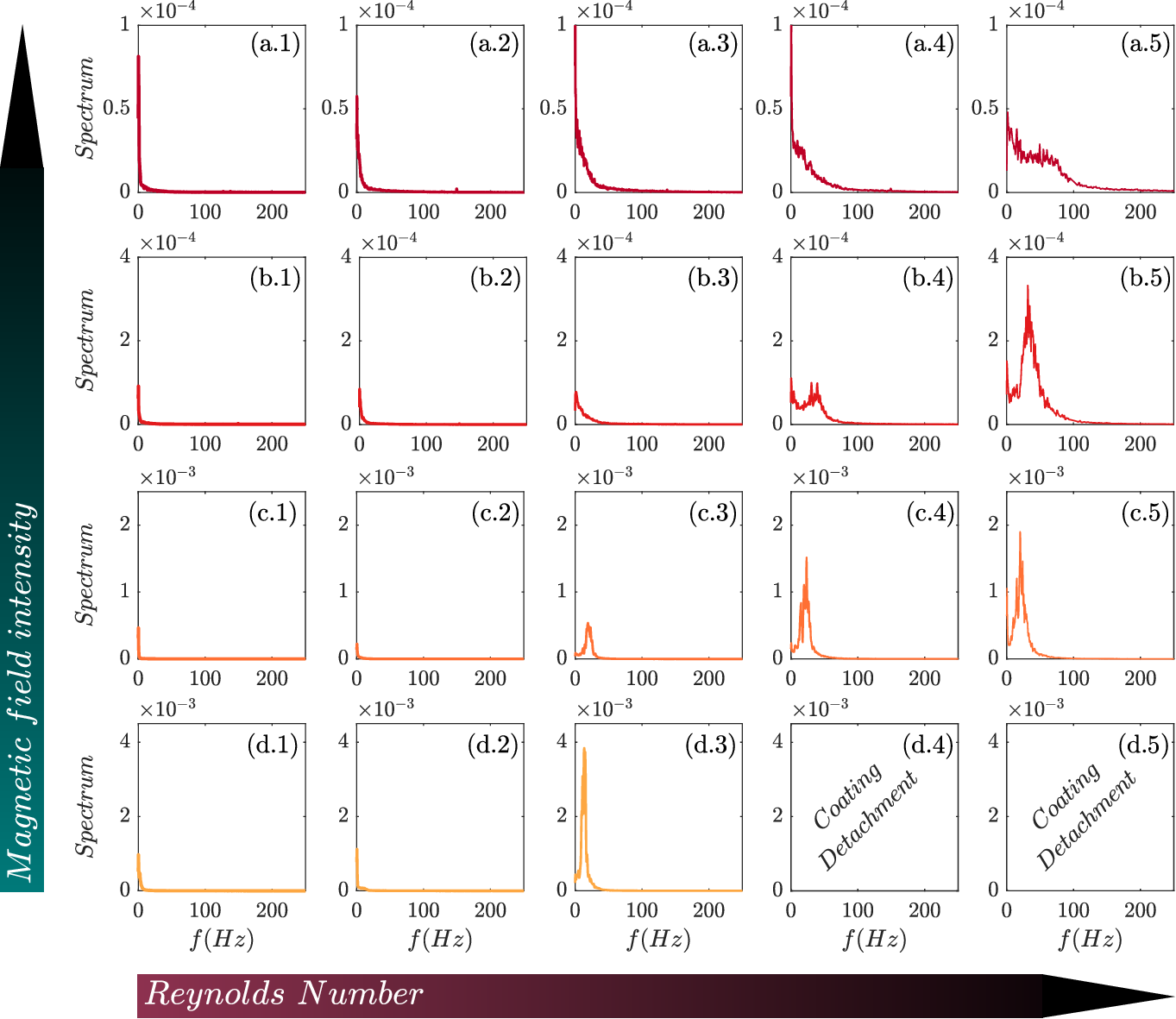}}
  \caption{Temporal power spectra of the interface elevation $\zeta$. The rows correspond to different magnetic field intensities, while the columns from left to right correspond to increasing Reynolds numbers.}
\label{fig:fig4}
\end{figure}

Figure \ref{fig:fig4} displays the temporal power spectrum density of the interface elevation for the experiments detailed in Table \ref{tab:tab1}. In this figure, the Reynolds number increases from left to right, while the magnetic field intensity ($H$) rises from the bottom to the top. Focusing on the first column a.1 - d.1, which pertains to experiments with the lowest Reynolds numbers, we observe that the spectra exhibit maxima near the origin of the abscissa, followed by a rapid decay as the frequency ($f$) increases. This suggests that, the interface remains mostly flat in these experiments.

Let us now consider the scenario with $H=460Gs$ (first row). As the Reynolds number increases, the power spectra begin to fill the range of $0 - 100 Hz$, with no distinct peak. That is, a range of waves starts to emerge without a dominant frequency. As the magnetic field decreases, a clear peak becomes evident in the power spectra at higher Reynolds numbers (as seen in, for instance, b.5, c.5, and d.3). The prominence of this peak intensifies as the flow cases move upward from the corresponding instability curves depicted in Figure \ref{fig:fig3}. The presence of a distinct peak suggests the occurrence of monochromatic cylindrical waves at the interface (refer to cases c.4, c.5, and d.3). In instances of multiple frequencies, the interface is characterized by a train of waves exhibiting different harmonics (see case a.5), indicating irregular wave patterns.  
Finally, increasing the magnetic field leads to a lower frequency manifestation of instability ($H$ acting as a stabilizing factor), indicated by a rightward shift in the peak (observe cases c.5 and b.5).
    
\begin{figure}
  \centerline{\includegraphics[scale=0.52]{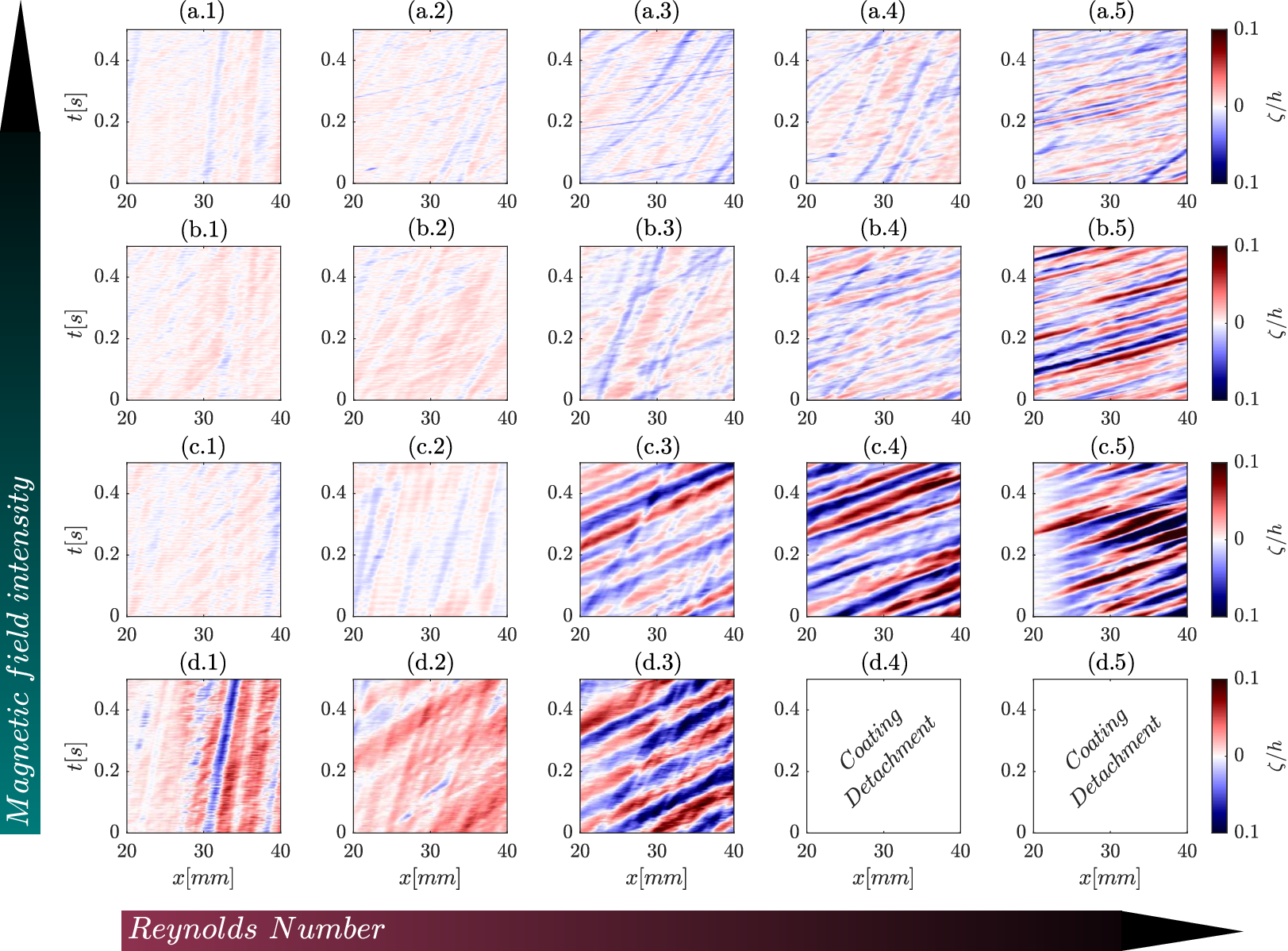}}
  \caption{Space–time diagrams of the interface elevation $\zeta$ for the same experiments shown in figure \ref{fig:fig4}.}
\label{fig:fig5}
\end{figure}

To characterize the spatio-temporal wave dynamics, in Figure \ref{fig:fig5} we present space-time diagrams of the interface elevation $\zeta$. Observing Figure \ref{fig:fig5}, it is noticeable that the interface displays a relatively flat profile when the Reynolds numbers are low and the magnetic field intensities are high, as depicted in the panels situated at the upper left corner of the figure.  However, as the Reynolds number increases and the magnetic field intensity decreases, clear wave crest-trough patterns become evident.

Additionally, in scenarios featuring intermediate Reynolds numbers and higher magnetic field strengths, exemplified by Figure \ref{fig:fig5}, e.g. a.3, the interface maintains a relatively flat profile. In contrast, when considering the same Reynolds number but with lower magnetic field intensity, as exemplified in Figure \ref{fig:fig5} d.3, significantly elevated interface profiles become apparent. It is worth noting that for flow cases displaying distinct peaks in the power spectrum diagrams (e.g. Figure \ref{fig:fig4} c.4, c.5, or d.3), Figure \ref{fig:fig5} shows a regular wave train (monochromatic waves). In cases dominated by a spectrum of frequencies, as seen in a.5, numerous harmonics emerge (colors alternate irregularly without a discernible trend) in the space–time diagrams. This observation corroborates our findings from the Fourier space analysis.

\subsection{Impact of ferrofluid interface stability on the flow}\label{subsec:Flow Measurements}

To correlate our observations regarding the characterization of the interface (Section \ref{subsec:Interface Characterization}) with drag reduction, we employ the velocity measurements obtained from 2D-PTV. For clarity, we focus on three distinct flow cases, namely a.5, b.5, and c.5 detailed in Table \ref{tab:tab1}, which correspond to the scenarios with the highest Reynolds number ($Re \approx 12500$) under varying magnetic field intensities: $H=460Gs$, $H=250Gs$, and $H=170Gs$.

In the subsequent part of this section, we normalize the vertical length scale ($y$) with respect to the channel fluid height ($D-h$) and mirror the profiles concerning the half-coated wall (depicted as colored lines) with those of the rigid wall side for each experiment (illustrated as black lines). As a result, the origin $y=0$ corresponds to both the ferrofluid interface (colored lines) and the rigid wall location (black lines).

\begin{figure}
  \centerline{\includegraphics[scale=0.65]{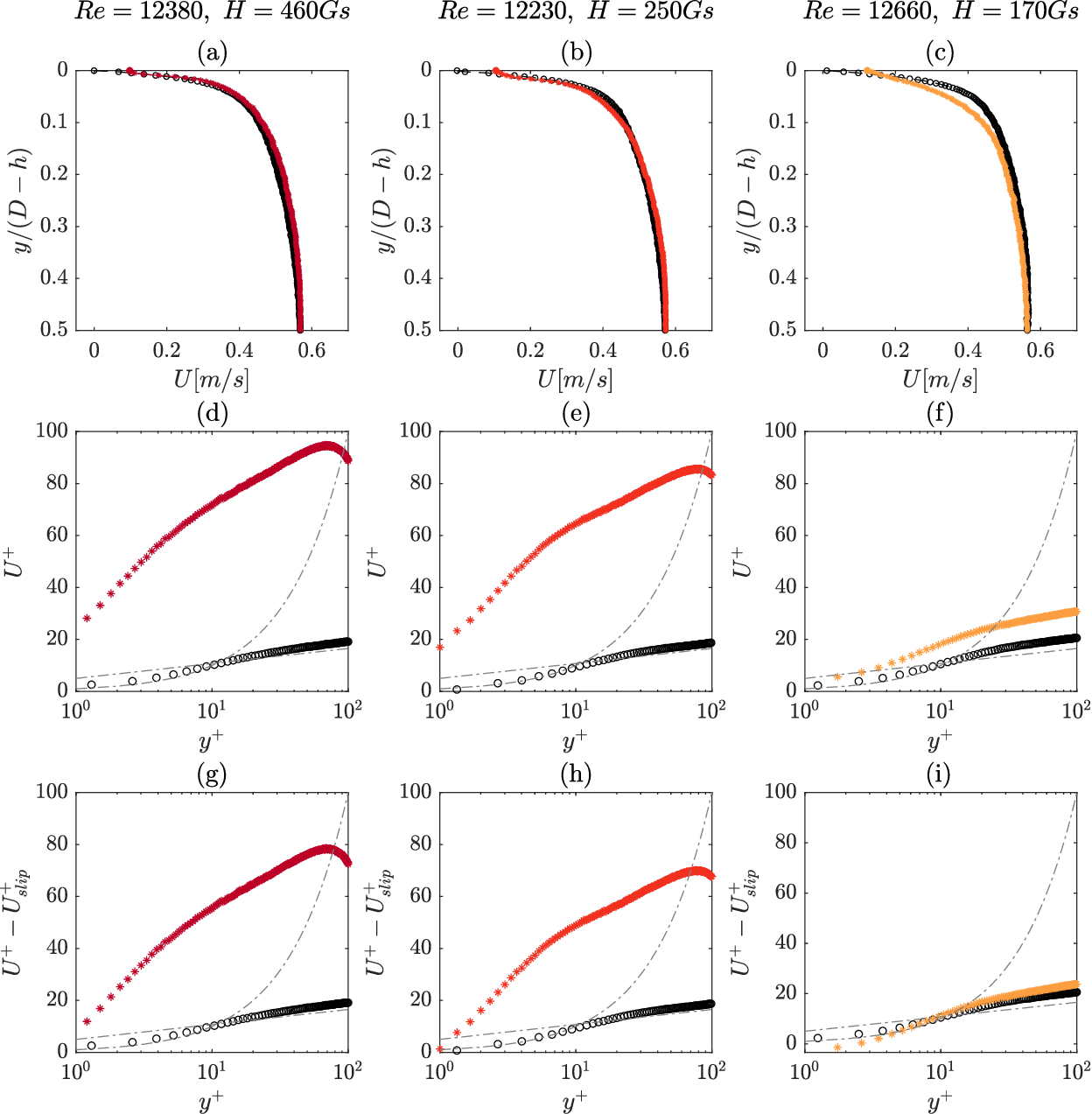}}
  \caption{Mean  streamwise velocity profiles of a.5, b.5 and c.5 (columns) in large scale units (a-c), wall-units (d-f) and wall-units with subtracted the slip velocity (g-i). For the color of the lines see Table \ref{tab:tab1}. In black the corresponding rigid wall profiles.}
\label{fig:fig6}
\end{figure}

The first row of Figure \ref{fig:fig6} illustrates the mean streamwise velocity profiles. We note a distinct slip velocity $U_{slip}$ at the ferrofluid interface for all three experiments, contrasting with the classical no-slip condition observed at the rigid wall. We also note differences away from the origin implying an asymmetry of the velocity profile in the bulk. While for the strongest magnetic field (Figure \ref{fig:fig6}a), the average velocity is always higher in the coated-side with respect to the rigid-wall one, for the highest magnetic field intensity experiments (Figure \ref{fig:fig6}c) the flow is generally faster in the lower half of the channel. In the case of $H=250Gs$ the "no-slip" flow is locally faster at $y/(D-h)\approx 0.05$. This difference in velocity profiles might be associated with a difference  of turbulence intensity in the interface proximity (see below).  

In the second row of Figure \ref{fig:fig6}, we show the average velocity profiles in wall units $y^{+} = y u_{\tau}/ \nu$ and $U^{+} = U/u_{\tau}$, with $u_{\tau} = (\tau_{0} / \rho)^{1/2}$ where $\tau_{0}$ is the space and time average wall shear stress for the rigid wall and the total shear stress $\tau_{c}$ at the ferrofluid interface, respectively. For the rigid-wall side, the wall shear stress is $\tau_{w} = \mu dU/dy|_{wall}$ evaluated at the wall location, while for the coated wall the total shear stress is computed as $\tau_{c} = \mu dU/dy - \rho <u'v'>$ at the interface location, where $\rho$ is the water density, $u'$ and $v'$ velocity fluctuations in the streamwise and wall-normal directions. Here, $<\cdot>$ denotes averaging over time and the streamwise direction $x$, given the statistical stationarity and homogeneity of the flow in $x$. Also depicted in the figure is the theoretical law of the wall, comprising the linear law $u^{+} = y^{+}$ of the viscous sub-layer and the logarithmic law $u^{+} = 1/\kappa\ ln(y^{+}) + C^{+}$ characterizing the log-law region (with $\kappa=0.4$ and $C^{+}=5$). As can be seen from second row of figure \ref{fig:fig6}, rigid-wall profiles cases are in good agreement with both laws, whereas a markedly distinct behavior is observed at the interface. In experiments a.5 and b.5 (Figure \ref{fig:fig6}d and e), there is a significant deviation between the experimental velocity profiles and the law of the wall, occurring in both the viscous sub-layer and the log-region. In contrast, experiment c.5 (Figure \ref{fig:fig6}f) is closer to the no-slip profile. As reported in existing literature \citep{sta2024phys} concerning an undulated boundary with slip velocity, the departure from the law of the wall within the viscous sub-layer primarily results from the slip, while the deviation in the log-layer is linked to wall waviness. To disentangle the two contributions, in the bottom row of Figure \ref{fig:fig6}, we present the same average velocity profiles in wall units after subtracting the contribution of the slip velocity, denoted as $U_{slip}$. As can be seen, a notable difference persists for flow cases a.5 and b.5 (Figure \ref{fig:fig6}g and h), whereas the profile of case c.5 nearly conforms to the no-slip profile (Figure \ref{fig:fig6}i). In essence, this implies that a significant part of the deviation for the c.5 case can be attributed to the slip velocity, while for the other two cases, the primary influence stems from interface waviness.

\begin{figure}
  \centerline{\includegraphics[scale=0.65]{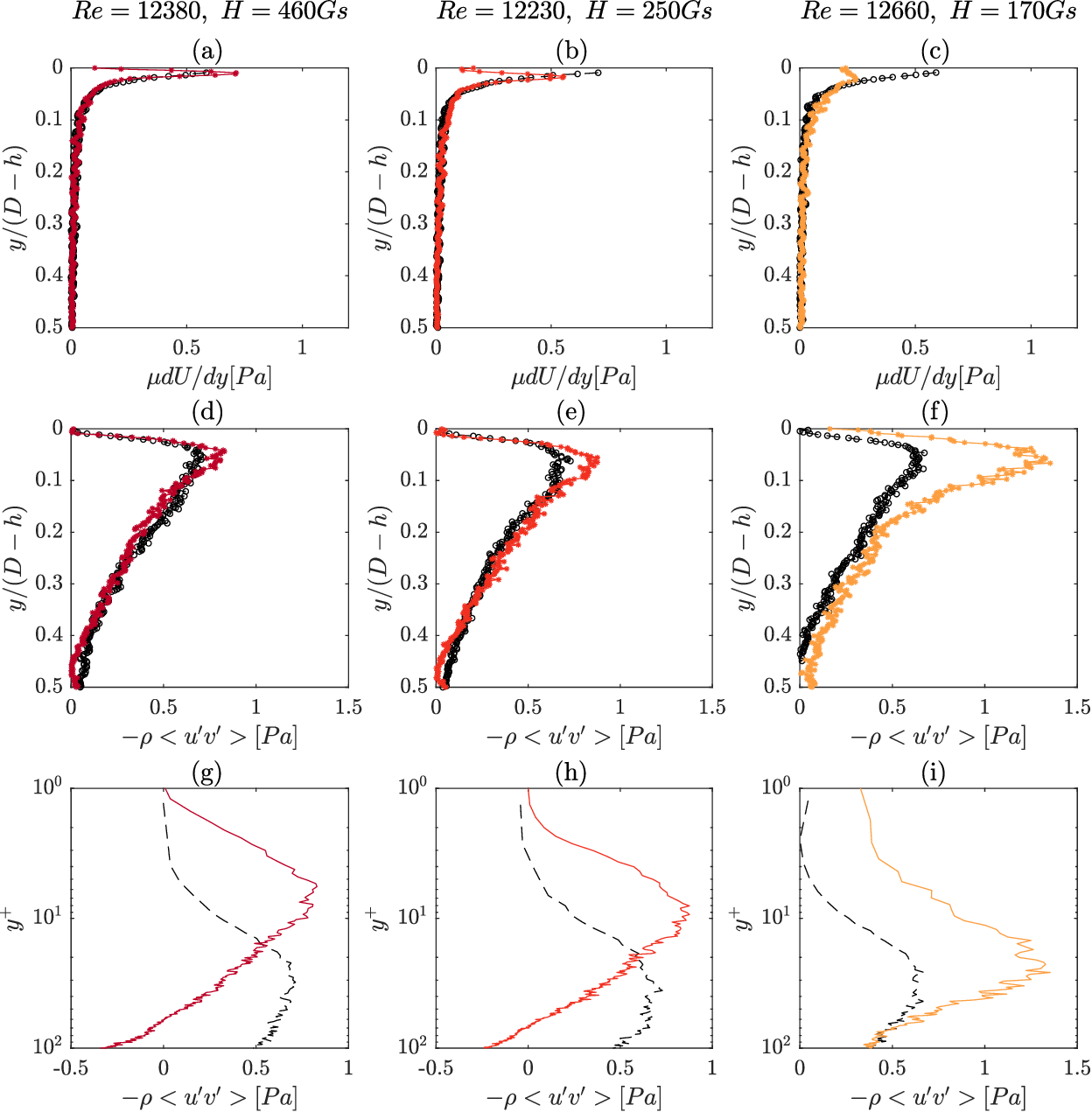}}
  \caption{Mean shear rate profiles (a-c) of a.5, b.5 and c.5 (columns) experiments, mean Reynolds shear stress profiles (d-f) and mean Reynolds shear stress profiles in wall  units (g-i). For the color of the lines see Table \ref{tab:tab1}. In black the corresponding rigid wall profiles.}
\label{fig:fig7}
\end{figure}

Previously, we noted that the ferrofluid layer introduces a slip condition at the upper wall of the channel, potentially affecting the shear rate in the channel flow. In Figure \ref{fig:fig7}, we further explore the influence of the ferrofluid layer on both viscous and turbulent shear rate profiles. In the top row of Figure \ref{fig:fig7}, we present the profiles of the viscous shear rate $\mu dU/dy$, where $\mu$ is the dynamic viscosity of the water and $U$ the space ($x$-direction) and time average velocity profile. As expected, the profiles for the rigid wall exhibit a peak at the wall location, whereas for the coated wall, the maximum value is attained in the vicinity of the interface. However, it sharply diminishes toward the interface, where a considerably smaller value is reached, owing to the slip condition. With a decrease in magnetic field intensity, the maximum value of the viscous shear rate also significantly diminishes. This reduction may be associated with a change in viscosity within the ferrofluid layer, which decreases with decreasing magnetic field intensity $H$ (\citet{sta2024phys}). 

In the second row of Figure \ref{fig:fig7}, we present the profiles of the turbulent shear stress $-\rho <u'v'>$. It is evident that, in the case of the most intense magnetic fields (d-f), the turbulent shear stress is comparable to that observed on the rigid wall side. In contrast, for lower magnetic intensities, $-\rho <u'v'>$ is notably higher especially around the peak location $y/(D-h)\approx 0.08$. This implies that additional velocity fluctuations are caused by waves, which were observed to have a higher amplitude $a$ (at $Re\approx const.$) for lower magnetic fields (Section \ref{subsec:Interface Characterization}). 

In the final row of Figure \ref{fig:fig7}, we present the Reynolds shear stresses in wall units. Notably, for cases with the most intense magnetic field, the peak is closer to the coated wall as compared to rigid wall (Figure \ref{fig:fig7}g-h). A different behavior is evident in cases of lower magnetic field (Figure \ref{fig:fig7}i), where the peak of $-\rho <u'v'>$ aligns with the rigid wall case. Interestingly, in this case, $-\rho <u'v'>$ remains high and positive in the bulk. This sustained non-zero $-\rho <u'v'>$ value evidences a detrimental impact of the interface instability, i.e., interface waviness causes an additional turbulent shear stress in close proximity to the coated wall.   

\begin{figure}
  \centerline{\includegraphics[scale=0.65]{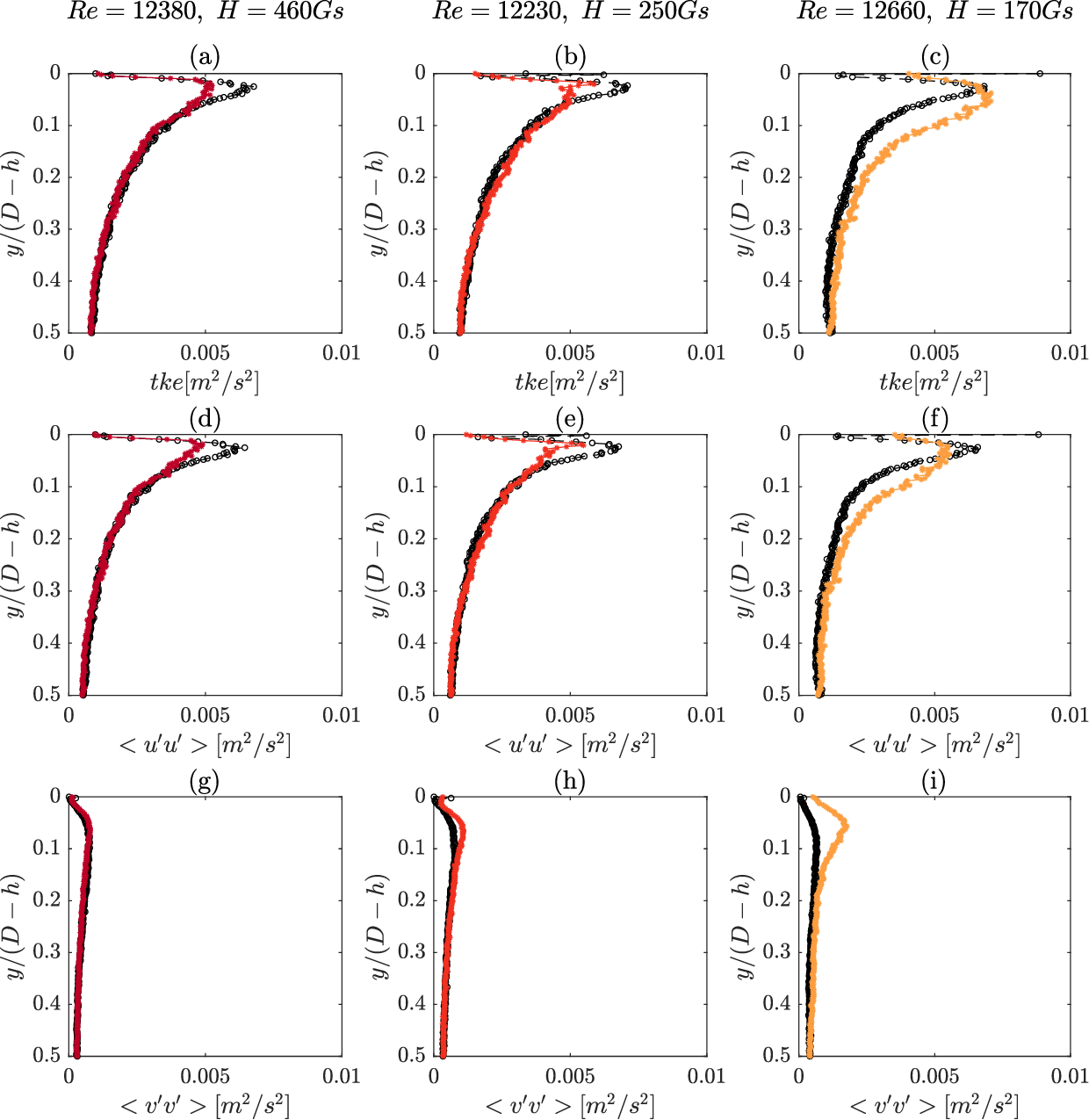}}
  \caption{Turbulent kinetic energy profiles (a-b) of a.5, b.5 and c.5 
 experiments (columns), stream-wise velocity (d-f) and wall normal (g-i) component contribution. For the color of the lines see Table \ref{tab:tab1}. In black the corresponding rigid wall profiles.}
\label{fig:fig8}
\end{figure}

To further characterize turbulence, in the first row of Figure \ref{fig:fig8}, we depict the profile of turbulent kinetic energy ($tke=<u'^2 + v'^2>$). Notably, $tke$ tends to be smaller in flow cases with higher $H$ (a-b) as compared with rigid wall side, while higher values can be observed for lower magnetic field intensity (Figure \ref{fig:fig8}c). To dissect the individual components, we present in the final two rows of Figure \ref{fig:fig8} the contributions of the streamwise and wall-normal components. The presence of the coating consistently diminishes the peak value of the streamwise component. However, in the case of lower magnetic field intensity, the wall-normal component experiences a significant increase. This is consistent with the increase in wave amplitude with decreasing $H$, as qualitatively noted in Figure \ref{fig:fig5}. In essence, the coating proves instrumental in reducing turbulent kinetic intensity up to a point where the waves generated at the interface reach an amplitude substantial enough to markedly elevate the wall-normal velocity fluctuation components.

\subsection{Drag reduction efficiency in relation to ferrofluid interface stability}\label{subsec:Drag Reduction}

\begin{figure}
  \centerline{\includegraphics[scale=0.65]{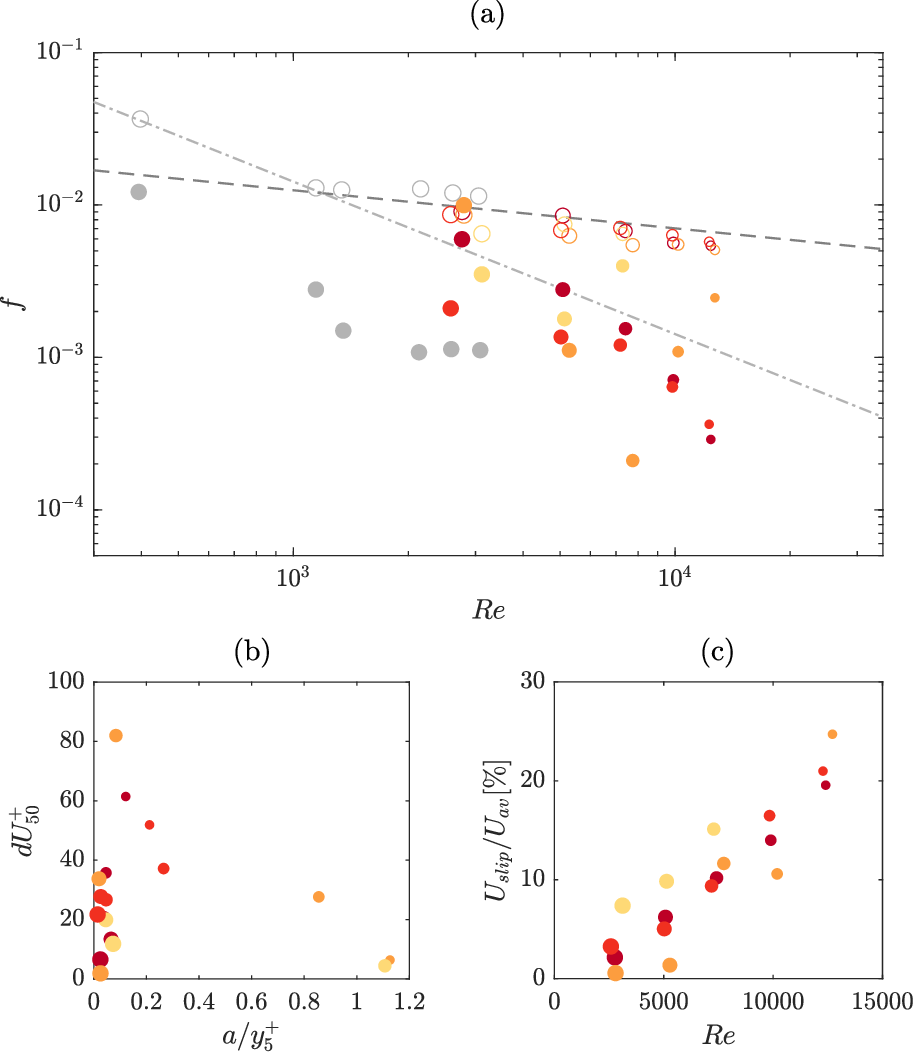}}
  \caption{(a) Friction factor $f$ against the Reynolds number $Re$. The filled and the open symbols represent respectively the results for the bottom and the top of the channel. For the colors of the symbols, refer to table \ref{tab:tab1}. Grey symbols represent data from \citet{sta2024phys}. Blu dashed line the solution for turbulent flows, while the green dashed line is the solution for laminar flows. (b) Difference between the experimental velocity profile and the theoretical log-law $dU_{50}^{+}$ at $y^{+}=50$, plotted against the normalized wave amplitude $a$ with respect to the viscous length scale $y^{+}=5$. (c) Ratio between the slip velocity $U_{slip}$ and the bulk average velocity $U_{av}$ against the Reynolds number $Re$.}
\label{fig:fig9}
\end{figure}

We calculate the Fanning friction factor, denoted as $f = \frac{2\tau_{0}}{\rho_w U^2}$, for both the bottom rigid wall and the interface. These results are illustrated in Figure \ref{fig:fig9}a. The figure also includes the theoretical solutions for laminar flows, represented by $f = \frac{c}{Re}$ with $c = 14.2$ \citep{kakacc1987handbook, hartnett1986hydrodynamic}, and turbulent flows, depicted as $f = 0.079(c_{d} \cdot Re)^{-1/4}$ with $c_{d} = 1.125$ \citep{choi1999thermal, jones1976improvement}. The figure incorporates friction factor data (depicted in grey) from \citet{sta2024phys} in which the authors used a similar setup to the present one. In those experiments, the channel had smaller dimensions, measuring $10\times 10 mm^2$ in cross-section, while an array of magnets directly affixed to the upper wall of the channel generated a magnetic field intensity of approximately $H=90Gs$ at the ferrofluid interface. Their experimental conditions encompassed both stable and unstable ferrofluid interface scenarios. Notably, due to the smaller cross-section in their setup, instability in the ferrofluid interface occurred at relatively low $Re$ numbers ($Re<4000$) compared to our current configuration, given the significantly higher bulk average velocity $U$ at the same $Re$.

From figure \ref{fig:fig9}a, it is evident that the friction factor values for the bottom wall (open symbols) align closely with the turbulent curve. In contrast, the majority of the values at the top interface are significantly below the turbulent curve, in some cases even below the laminar curve. That is, the drag experienced at the interface between the water and ferrofluid is generally much lower compared to that at the bottom of the channel. 

Upon closer examination of the coated wall cases, Figure \ref{fig:fig9}a illustrates a consistent trend where the friction factor decreases with increasing Reynolds number ($Re$), particularly evident in experiments conducted with higher magnetic field intensities, namely $H=460Gs$ and $H=250Gs$. This holds true also for experiments with lower magnetic field intensities ($H=170Gs$ and $H=140Gs$), except for cases at higher Reynolds number where an inversion in this trend occurs (refer to experiments c.4, c.5, and d.3). In these cases, the friction factor increases with $Re$ at a given $H$, albeit remaining below the values observed for rigid wall. A noteworthy characteristic of experiments c.4, c.5, and d.3 is the high wave amplitudes formed at the interface of the ferrofluid layer compared to other experiments (refer to Figure \ref{fig:fig5}). This may be linked to an increase of the turbulence intensity in the interface proximity due to additional velocity fluctuations connected to the interface instability. Generally, the slip condition and interface waves contribute to drag reduction. However, when the waves reach high amplitudes, they appear to transition from being advantageous to becoming detrimental for drag reduction. This notion gains additional support through a direct comparison of our experimental data with those presented by \citet{sta2024phys} (depicted as grey symbols). Although there is a general consistency in trends between the two datasets, \citet{sta2024phys} noted smaller friction factors at smaller Reynolds numbers ($2000<Re<4000$) along the ferrofluid interface. This difference could stem from the noted earlier instability of the interface at these Reynolds numbers in their experiments, which had a beneficial impact on drag reduction. In contrast, in the current scenario, the interface remains flat at these Reynolds numbers, with the sole contributing factor being the slip condition. 

To analyze a possible relationship between drag reduction and the kinematics of the interface, we illustrate in Figure \ref{fig:fig9}b the disparity between the experimental velocity profile with subtracted slip velocity (see e.g. Figure \ref{fig:fig6}g-i) and the log-law $dU_{50}^{+}$ at $y^{+}=50$, plotted against the wave amplitude $a$ normalized by a viscous length scale, namely $y^{+}=5$. The choice of this viscous length scale is motivated by the fact that the viscous shear stress dominates near the ferrofluid interface (see e.g. Fig \ref{fig:fig8}a-b and d-e), i.e. even though there is a slip condition there is a viscous layer at the interface in analogy to the viscous sublayer near the solid wall. In figure \ref{fig:fig9}b, $dU_{50}^{+}$ quantifies the effect of the wall waviness on the shear stress. Indeed high positive values indicate a lower shear stress at the interface location, signifying drag reduction. Across all experiments, an increase in wave amplitudes corresponds to an increase in $dU_{50}^{+}$ as long as $a$ remains small compared to the viscous length scale. However, when the wave amplitudes approach the viscous length scale (experiments c.4, c.5, and d.3), the advantage derived from surface waviness diminishes significantly. This observation further substantiates our hypothesis that the positive effect of the waves on drag reduction diminishes with wave amplitude and may even become detrimental.

Finally, to complete our analysis of the observed frictional behavior at the coated wall, we present in Figure \ref{fig:fig9}c the percentage of slip velocity ($U_{slip}$) relative to the average bulk velocity ($U_{av}$). Notably, $U_{slip}/U_{av}$ exhibits nearly linear growth with increasing Reynolds numbers in the range covered by our experiments. This is consistent with the Reynolds number trend observed in the friction factor (Figure \ref{fig:fig9}a), where, in general, there is a decrease in $f$ with $Re$, except for cases c.4, c.5, and d.3. Nonetheless, these latter cases are characterized by an elevated slip velocity (Figure \ref{fig:fig9}c), mitigating to some degree the effects of extra velocity fluctuations brought about by the ferrofluid interface waves.

\begin{figure}
  \centerline{\includegraphics[scale=0.65]{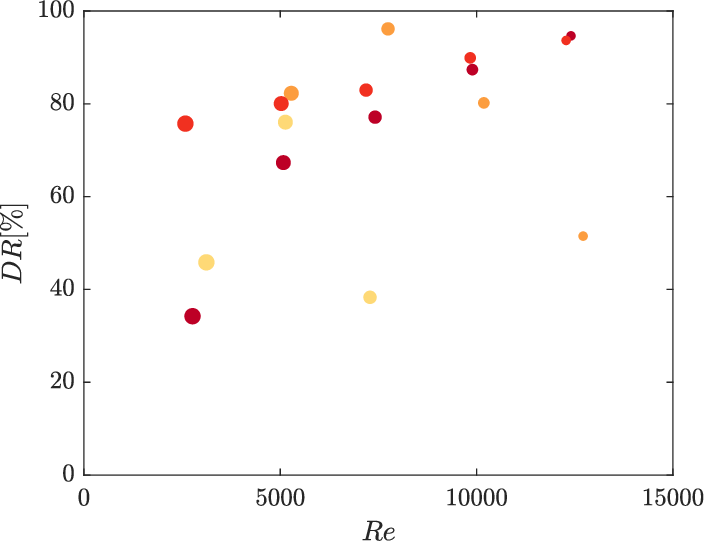}}
  \caption{Drag reduction $DR[\%]$ against the Reynolds number $Re$. The color-coding and symbol size are as specified in Table \ref{tab:tab1}. Not shown in the figure, experiment c.1 in which the drag coefficient is slightly negative (drag increase).}
\label{fig:fig10}
\end{figure}

A final element in assessing the effectiveness of the drag reduction technique is the so-called drag reduction coefficient. In Figure \ref{fig:fig10}, this coefficient, expressed as $DR[\%] = (\tau_{w} - \tau_{c})/\tau_{w}\cdot 100$, is depicted against the Reynolds number $Re$, with $\tau_{w}$ denoting the wall shear stress at the rigid wall and $\tau_{c}$ representing the corresponding value at the ferrofluid interface. It is evident that, in general, drag reduction tends to increase with the Reynolds number for a given $H$, except in cases where the interface wave amplitude is notably elevated compared to the viscous length scale $y^{+}_{5}$ (flow-cases c.4, c.5, and c.5). It is, however, noteworthy that remarkably high values of $DR$, such as $95\%$, can be achieved, exemplified in scenarios characterized by $Re \approx 12500$ and $H=460$ or $H=250$.

\section{Conclusions and summary}\label{sec:Conclusions}
This study addresses the application of a ferrofluid layer coating to reduce drag in turbulent channel flow with focus on the effect of ferrofluid interface stability. The ferrofluid layer is held in position by permanent magnets, creating a magnetic field oriented tangentially to the interface between the magnetic and diamagnetic fluids. To realize the channel flow a new experimental setup was devised, while a 2D-PTV and interface identification techniques were employed to capture the flow field and accurately locate the ferrofluid interface. 

Firstly, a stability analysis reveals that in experiments with higher flow rates, the interface between the ferrofluid layer and water experiences instability. This instability is confirmed by both a power spectral analysis and height-time diagrams. It was observed that as the Reynolds number increases or the magnetic field intensity decreases, instabilities occur, characterized by the formation of travelling waves on the interface. The amplitude of these waves increases with increasing $Re$ or decreasing $H$. These findings are consistent with the stability theory, further supporting the relationship between forcing, magnetic field intensity, and interface stability.

Subsequently, the ferrofluid layer effect on the velocity profile within the channel is investigated. An evident slip velocity emerges at the interface between the two fluids, introducing an asymmetry between the coated and rigid wall sides extending into the bulk of the flow. In experiments with a strong magnetic field, the flow is faster on the coated side compared to the rigid wall side, while the opposite occurs with a lower magnetic field. We inferred that this may be linked to increased shear stresses due to the increasing Reynolds stress with interface wave amplitude. To assess the individual contributions of slip and waviness to the modification of the velocity profile on the coated side, we scrutinize the velocity profiles in wall units. Our findings reveal that at high $H$, both slip velocity and wall waviness contribute to reduce wall shear stress, leading to an elevation of the velocity profile. Conversely, at low $H$, waviness induces extra turbulence. However, the net effect of the ferrofluid interface is still to reduce drag because of the dominant contribution of the slip velocity.

We explored the impact of the ferrofluid layer on both viscous and turbulent shear stresses within the channel. Notably, the viscous stress experiences a significant reduction at the interface compared to its solid wall counterpart, primarily attributable to the slip condition between the two fluids. Conversely, Reynolds shear stresses are elevated in the vicinity of the interface compared to the solid wall, yet they strongly decrease without vanishing at the interface location. 

The picture emerging from these observations is as follows: turbulent shear stresses increase near the interface compared to the solid wall when the interface becomes unstable, generating travelling wave patterns. However, right at the interface location, turbulent shear stresses are suppressed due to the non-penetrative condition between the two fluids. Despite the shear stress at the ferrofluid location having an additional turbulent contribution compared to the solid wall, this augmentation does not compensate for the viscous decrease by the slip velocity, resulting in a lower total shear stress for the coated wall compared to the rigid boundary. As the magnetic field intensity decreases, the Reynolds shear stresses are observed to increase, aligning with the increased turbulence intensity due to larger interface wave amplitudes in this scenario. An analysis of turbulence intensity corroborates this observation, highlighting that the primary contributor to the increase in turbulence intensity with decreasing $H$ is the augmentation in vertical velocity fluctuations.

The impact of the ferrofluid coating on drag reduction is assessed by examining the friction factor. The results demonstrate that, for most of the flow conditions investigated in this study, the friction factor at the location of the ferrofluid is significantly lower compared to the solid wall. Interestingly, this reduction in friction factor persists even when the interface between the two fluids experiences instability, leading to the formation of travelling waves. However, the amplitude of these waves proves crucial in determining whether wall waviness positively or negatively affects drag. Specifically, when the wave amplitude is significantly smaller than the viscous length scale, it has a favorable effect on drag reduction. Conversely, when the amplitude becomes comparable to the viscous length, it it becomes detrimental.


\backsection[Acknowledgements]{We are grateful for financial support from SNSF grant number 200727.}

\backsection[Declaration of interests]{The authors report no conflict of interest.}

\backsection[Author ORCIDs]{ M.M. Neamtu-Halic, https://orcid.org/0000-0002-6438-7967; M. Holzner, https://orcid.org/0000-0003-2702-8612; L.M. Stancanelli, https://orcid.org/0000-0001-9725-9964.}

\bibliographystyle{jfm}

\end{document}